\documentstyle[epsf]{aipproc}

\newcommand{\be}{\begin{eqnarray}}

\newcommand{\ee}{\end{eqnarray}}

\title{
        \begin{flushright}
        {\normalsize
        BNL-NT-99/4\\
        October 1999 \\}
        \end{flushright}
\bf     Two Lectures on Small x and High Gluon
Density\footnote{
Lectures XXXIX'th Crakow School of Theoretical Phyics, Zakopane Poland,
May 29 - June 8, 1999}  
       }
\author{Larry McLerran\\
       {\small\it Nuclear Theory Group,  Brookhaven National Laboratory
        Upton, NY 11973  } \\      
       }

\date{}

\begin{document}
\maketitle

Abstract:  In these lectures,  I shall discuss small x physics and the
consequences of the high gluon density which arises as x decreases.
I argue that an understanding of this problem would lead to knowledge
of the high energy asymptotics of hadronic processes.  The high gluon
density should allow a first principles computation of these asymptotics
from QCD.  This physics might be experimentally probed in heavy ion
colliders or in high energy electron-nuclear collisions

\section{Lecture I:  Lots of Problems}

\subsection{Introduction}

I think we all believe that QCD describes hadronic physics.  It has been
tested in a variety of environments.  For high energy short distance
phenomena, perturbative QCD computations successfully confront
experiment.  In lattice Monte-Carlo computations, one gets a successful
semi-quantitative description of hadronic spectra, and perhaps in the
not too distant future one will obtain precise quantitative agreement.

At present, however, all analytic computations and all precise QCD tests
are limited to the small class of problems which correspond to short
distance physics.  Here there is some characteristic energy transfer 
scale $E$, and one uses asymptotic freedom,
\be
        \alpha_S(E) \rightarrow 0
\ee
as $E \rightarrow \infty$

One question which we might ask is whether there are any
non-perturbative ``simple phenomena'' which arise from QCD which are
worthy of further effort.  The questions  I would ask before I would
become interested in understanding such phenomena are
\begin{itemize}

\item  Is the phenomenon simple and pervasive?

\item  Is it reasonably plausible that one can understand the phenomena
from first principles, and  compute how it would appear in nature?

\end{itemize}

I will in this lecture try to explain a wide class of phenomena in QCD
which are pervasive, and appear to follow simple patterns.  I will then
try to explain why I believe that these phenomena can be simply
understood within QCD.

\subsection{Total Cross Sections at Asymptotic Energy}

Computing total cross section as $E \rightarrow \infty$ is one of the
great unsolved problems of QCD.  
Unlike for processes which are computed in perturbation theory,
it is not required that any energy transfer become large as the total 
collision energy $E \rightarrow \infty$.  Computing a total cross section for 
hadronic scattering therefore appears to be intrinsically non-perturbative.
In the 60's and early 70's, Regge
theory was extensively developed in an attempt to understand the total
cross section.  The results of this analysis were to my mind
inconclusive, and certainly can not be claimed to be a first principles
understanding from QCD.

Typically, it is assumed that the total hadronic  cross section grows as
$ln^2
E$ as $E \rightarrow \infty$.  This is the so called Froisart bound
which corresponds to the maximal growth allowed by unitarity of the S
matrix.    Is this correct?  Is the coefficient of $ln^2 E$ universal
for all hadronic precesses?  Why is the unitarity limit saturated?  Can
we understand the total cross section from first principles in QCD?  Is
it understandable in weakly coupled QCD, or is it an intrinsically
non-perturbative phenomenon?

\subsection{How Are Particle Produced in High Energy Collisions?}

The total multiplicity of produced particles is an increasing function
of energy for hadronic collisions.  The total multiplicity for
$pp$ and for $\overline p p$ collisions has roughly the same energy
dependence.  Is whatever 
is causing the increase in multiplicity in these collisions arising from
the same mechanism?

The obvious question is can we compute $N(E)$, the total multiplicity of
produced particles as a function of energy?

At this point it is useful to develop some mathematical tools.  I will
introduce some useful kinematic variables:  light cone coordinates.
Let the light cone longitudinal momenta be
\be
        p^\pm = {1 \over \sqrt{2}} (E \pm p_z)
\ee
Note that the invariant dot product
\be
        p \cdot q = p_t \cdot q_t - p^+q^- - p^- q^+
\ee
and that
\be
        p^+p^- = {1 \over 2} (E^2 - p_z^2) = {1 \over 2} (p_T^2 +m^2) = {1
\over 2} m_T^2 \ee This equation defines the transverse mass $m_T$. 
(Please note that my metric is the negative of that conventionally used in
particle physics.  An unfortunate consequence of my education.  Students,
please feel free to convert everything to your favorite metric.)

Consider a collision in the center of mass frame.
The right moving particle has $p_1^+ \sim \sqrt{2} \mid p_z \mid$ and
$p_1^- \sim {1 \over {2\sqrt{2}}} m_T^2/\mid p_z \mid$.
For the colliding particles $m_T = m_{projectile}$, that is because the
transverse momentum is zero, the transverse mass equals the particle
mass For particle 2,
we have $p^+_2 = p^-_1$ and $p^-_2 = p^+_1$.

If we define the Feynman $x$ of a produced pion as
\be
        x = p^+_\pi /p_1^+
\ee
then $0 \le x \le 1$.  The rapidity of a pion is defined to be
\be
        y = {1 \over 2} ln(p^+_\pi /p^-_\pi) = {1 \over 2} ln(2p^{+2}/m_T^2)
\ee
For pions, the transverse mass includes the transverse momentum of the
pion.

The pion rapidity is always in the range $-y_{CM} \le y \le y_{CM}$ where
$y_{CM} = ln({p^+/m_{projectile}})$  All the pions are produced in a
distribution of rapidities within this range.

These definitions are useful, among other reasons, because of their simple
properties under longitudinal Lorentz boosts:  $p^\pm \rightarrow
\kappa^{\pm 1} p^{\pm}$ where $\kappa$ is a constant.  Under boosts,
the rapidity just changes by a constant.  
(Students, please check this relationship for momenta under boosts.)

It is convenient in
the center of mass frame to think of the positive rapidity pions as
somehow related to the right moving particle and the negative rapidity
particles as related to the left moving particles.  We define
$x = p^+/p^+_{projectile}$ and $x^\prime = p^-/p^-_{projectile}$
and use $x$ for positive rapidity pions and $x^\prime$ for negative
rapidity pions.

Of course more than just pions are produced in high energy collisions.
The variables we just presented easily generalize to these particles.

Several theoretical issues arise in multiparticle production.  Can we
compute $dN/dy$?  or even $dN/dy$ at $y = 0$?  How does the average transverse
momentum of produced particles $<p_T>$ behave with energy?  What is the
ratio of produced strange/nonstrange, and corresponding rations of
charm, top, bottom etc at $y = 0$ as the center of mass energy
approaches infinity?

Does multiparticle production as $E \rightarrow \infty$ at
$y = 0$ become
\begin{itemize}

\item Simple?

\item Understandable?

\item Computable?

\end{itemize}

\subsection{Deep Inelastic Scattering}

In deep inelastic scattering, an
electron emits a virtual photon which scatters 
from a quark in a hadron.  The momentum and energy
transfer of the electron is measured, and the results of the break up
are not.  In these lectures, we cannot develop the theory of deep
inelastic scattering.  Suffice it to say, that this measurement is
sufficient at large momenta transfer $Q^2$ to measure the distributions
of quarks in a hadron.

To describe the quark distributions, it is convenient to work in a
reference frame where the hadron has a large longitudinal momentum
$p^+_{hadron}$.  The corresponding light cone momentum of the
constituent is $p^+_{constituent}$.  We define $x =
p^+_{constituent}/p^+_{hadron}$.  (This x variable is equal to the
Bjorken x variable, which can be defined in a frame independent way.
In this frame independent definition, $x = Q^2/2p\cdot Q$ where $p$ is 
the momentum of the hadronic target and $Q$ is the momentum of the virtual
photon.  Students, please check that this is true.)
The cross section which one extracts in deep inelastic scattering can be
related to the distributions of quarks inside a hadron, $dN/dx$.

It is useful to think about the distributions as a function of rapidity.
We define this for deep inelastic scattering as
\be
        y = y_{hadron} - ln(1/x)
\ee
and the invariant rapidity distribution as
\be
        dN/dy = x dN/dx
\ee

A  $dN/dy$ distribution for constituent gluons
is similar to that for the distribution of produced pions in hadronic
collisions. The main difference is that in deep inelastic scattering, the 
rapidity goes from that of the projectile to zero, while in hadron hadron 
scattering, it goes from the rapidity of the projectile to that of the
target.  In the center of mass frame for hadron collisions,
the plot is essentially twice as wide as that for deep inelastic
scattering, and looks roughly like the deep inelastic distribution folded
over on itself.

We shall later argue that there is in fact a relationship between the
structure functions as measured in deep inelastic scattering and the
rapidity distributions for particle production.  We will argue that the
gluon distribution function is in fact proportional to the pion rapidity
distribution.

The small x problem is that in experiments at Hera, the rapidity
distribution function for quarks grows as the rapidity difference between
the quark and the hadron grows.  This growth appears to be more rapid
than simply $\mid y_{proj} - y \mid$ or $(y_{proj}-y)^2$, 
and various theoretical models based on the
original considerations of Lipatov and colleagues suggest it may grow as
an exponential in $\mid y_{proj} - y \mid$.\cite{klf}  
If the rapidity distribution grew at most as
$y^2$, then there would be no small x problem.
We shall try to explain
the reasons for this later in this lecture.

At HERA, the gluon structure function has been measured at
various values of $Q^2$ and $x$..\cite{z}
The gluon distribution $xg(x)$  rises at small x. This is the small x
problem.  It is amusing that the rise in the gluon structure function 
at small x is similar qualitatively to the increase in the pion
multiplicity in high energy hadronic interactions.

Why is the small x rise in the gluon distribution a problem?\cite{glr}
Imagine we view a nucleon head on.  
As we add more and more 
constituents, the hadron becomes
more and more crowded.  If we were to try to measure these constituents
with say an elementary photon probe, as we do in deep inelastic
scattering, we might expect that the hadron would become so crowded that
we could not ignore the shadowing effects of constituents as we make the
measurement.  (Shadowing means that some of the partons are
obscured by virtue of having another parton in front of them.  For hard spheres,
for example, this would result in a decrease of the 
scattering cross section relative to what is
expected from incoherent independent scattering.)

In fact, in deep inelastic scattering, we are measuring the cross section
for a virtual photon $\gamma^*$ and a hadron, $\sigma_{\gamma^*
hadron}$.
Making x smaller correspond to increasing the energy of the interaction
(at fixed $Q^2$).  An exponential growth in the rapidity corresponds to
power law growth in $1/x$, which in turn implies power law growth with
energy.  This growth, if it continues forever, violates unitarity.  The
Froissart bound will allow at most $ ln^2 (1/x)$.  (The Froissart
bound is a limit on how rapidly a total cross section can rise.  It
follows from the unitarity of the scattering matrix.)

We shall later argue that in fact the distribution functions at fixed
$Q^2$ do in fact saturate and cease growing so rapidly at high energy.
The total number of gluons however demands a resolution scale, and we
will see that the natural intrinsic scale is growing at smaller values
of x, so that effectively, the total number of gluons within this
intrinsic scale is always increasing.  The quantity
\be
        \Lambda^2 = {1 \over {\pi R^2}} {{dN} \over {dy}}
\ee
defines this intrinsic scale.  
Here $\pi R^2$ is the cross section for hadronic scattering from the hadron.
For a nucleus, this is well defined.  For a hadron, this is less certain,
but certainly if the wavelengths of probes are small compared to $R$, this should
be well defined.
If
\be
        \Lambda^2  >> \Lambda^2_{QCD}
\ee
as the Hera data suggests, then we are dealing with weakly coupled QCD
since $\alpha_S(\Lambda) << 1$.

Even though QCD may be weakly coupled at small x, that does not mean the
physics is perturbative.  There are many examples of nonperturbative
physics at weak coupling.  An example is instantons in electroweak
theory, which lead to the violation of baryon number.   Another example
is the atomic physics of highly charged nuclei.  The electron propagates
in the background of a strong nuclear Coulomb field, but on the other
hand, the theory is weakly coupled and there is a systematic weak
coupling expansion which allows for systematic computation of the
properties of high Z (Z is the charge of the nucleus) atoms.

If the theory is local in rapidity, then the only parameter which can
determine the physics at that rapidity is $\Lambda^2$.
(Locality in rapidity means that there are not long range correlations
in the hadronic wavefunction as a function of rapidity.  In pion production,
it is known that except for overall global conserved quantities such as 
energy and total charge, such 
correlations are of short range.)
Note that if only $\Lambda^2$ determines the physics, then in an
approximately scale invariant theory such as QCD,  a typical transverse
momentum of a constituent will also be of order $\Lambda^2$.  If
$\Lambda^2 >> 1/R^2$, where $R$ is the radius of the hadron, then
the finite size of the hadron becomes irrelevant.  Therefore at small
enough x, all hadrons become the same.  The physics should only be
controlled by $\Lambda^2$.

There should therefore be some equivalence between nuclei and say
protons.  When their $\Lambda^2$ values are the same, their physics
should be the same.  We can take an empirical parameterization of the
gluon structure functions as
\be
        {1 \over {\pi R^2}} {{dN} \over {dy}} \sim {{A^{1/3}} \over x^\delta}
\ee
where $\delta \sim .2 - .3$.  This suggests that there should be the
following correspondences:
\begin{itemize}

\item RHIC with nuclei $\sim$ Hera with protons

\item LHC with nuclei $\sim$ Hera with nuclei

\end{itemize}

Since the physics of high gluon density is weak coupling we have the
hope that we might be able to do a first principle calculation of
\begin{itemize}

\item the gluon distribution function

\item the quark and heavy quark distribution functions

\item the intrinsic $p_T$ distributions quarks and gluons

\end{itemize}

We can also suggest a simple escape from unitarity arguments which
suggest that the gluon distribution function must not grow at
arbitrarily small x.  The point is that at smaller x, we have larger
$\Lambda$ and correspondingly larger $p_T$.  A typical parton added to
the hadron has a size of order $1/p_T$.  Therefore although we are
increasing the number of gluons, we do it by adding in more gluons of
smaller and smaller size.  A probes of size resolution $\Delta x \ge
1/p_T$  at fixed $Q$ will not see partons smaller than this resolution
size.  They therefore do not contribute to the fixed $Q^2$ cross
section, and there is no contradiction with unitarity.

\subsection{Heavy Ion Collisions}

Imagine we have two Lorentz contracted
nuclei approaching one another at the speed of light.  Since they are
well localized, they can be thought of as sitting at $x^\pm = 0$, that
is along the light cone, for $t < 0$  At $x^\pm = 0$, the nuclei
collide.  To analyze this problem for $t \ge 0$, it is convenient to
introduce a time variable which is Lorentz covariant under longitudinal
boosts
\be
        \tau = \sqrt{t^2 - z^2}
\ee
and a space-time rapidity variable
\be
        \eta = {1 \over 2} ln\left( {{t-z} \over {t+z}} \right)
\ee
For free streaming particles
\be
        z = vt = {p_z \over E} t
\ee
we see that the space-time rapidity equals the momentum space rapidity
\be
        \eta = y
\ee

If we have distributions of particles which are slowly varying in
rapidity, it should be a good approximation to take the distributions to
be rapidity invariant.  This should be valid at very high energies in
the central region.  By the correspondence between space-time and
momentum space rapidity, it is plausible therefore to assume  that
distributions are independent of $\eta$.  Therefore distributions are
the same on lines of constant $\tau$.
At $z = 0$, $\tau = t$, so that $\tau $ is a longitudinally Lorentz
invariant time variable.

We expect that at very late times, we have a free streaming gas of
hadrons.  These are the hadrons which eventually arrive at our detector.
At some earlier time, these particle decouple from a dense gas of
strongly interacting hadrons.  As we proceed earlier in time, at some
time there is a transition between a gas of hadrons and a plasma of
quarks and gluons.  This may be through a first order phase transition
where the system might exist in a mixed phase for some length of time,
or perhaps there is a continuous change in the properties of the system

At some earlier time, the quarks and gluons of the quark-gluon plasma
are formed.  This is at some time of the order of a Fermi, perhaps as
small as $.1 ~Fermi$.  As they form, the particles scatter from one
another, and this can be described using the methods of transport
theory.  At some later time they have thermalized, and the system can be
approximately described using the methods of perfect fluid
hydrodynamics.

In the time between that for which the quarks and gluons have been
formed and $\tau = 0$, the particles are being formed.  This is where
the initial conditions are made.

In various levels of sophistication, one can compute the properties of
matter made in heavy ion collisions at times later than the formation
time.  The problems are understood in principle for $\tau \ge
\tau_{formation}$ if perhaps not in fact.
Very little is known about the initial conditions.

In principal, understanding the initial conditions should be the
simplest part of the problem.  At the initial time, the degrees of
freedom are most energetic and therefore one has the best chance to
understand them using weak coupling methods in QCD.

There are two separate classes of problems one has to understand for the
initial conditions.  First the two nuclei which are colliding are in
single quantum mechanical states.  Therefore for some early time, the
degrees of freedom must be quantum mechanical.  This means that
\be
        \Delta z \Delta p_z \ge 1
\ee
Therefore classical transport theory cannot describe the particle down
to $\tau = 0$ since classical transport theory assumes we know a
distribution function $f(\vec{p}, \vec{x}, t)$, which is a simultaneous
function of momenta and coordinates.  This can also be understood
as a consequence of entropy.  An initial quantum state has zero entropy.
Once one describes things by classical distribution functions,
entropy has been produced.  Where did it come from?

Another problem which must be understood is classical charge coherence.
At very early time, we have a tremendously large number of particles
packed into a longitudinal size scale of less than a fermi.  This is due
to the Lorentz contraction of the nuclei.  We know that the particles
cannot interact incoherently.  For example, if we measure the field due
to two opposite charge at a distance scale $r$ large compared to their
separation, we know the field fall as $1/r^2$, not $1/r$.  On the other
hand, in cascade theory, interactions are taken into account by cross
sections which involve matrix elements squared.  There is no room for
classical charge coherence.

There are a whole variety of problems one can address in heavy ion
collisions such
\begin{itemize}

\item What is the equation of state of strongly interacting matter?

\item Is there a first order QCD phase transition?

\end{itemize}
These issues and others would take us beyond the scope of these
lectures.  The issues which I would like to address are related to the
determination of the initial conditions, a problem which can hopefully
be addressed using weak coupling methods in QCD.

\subsection{Universality}

There are two separate formulations of universality which are important
in understanding small x physics.

The first is a weak universality.  This is the statement that physics
should only depend upon the variable
\be
        \Lambda^2 = {1 \over {\pi R^2}} {{dN} \over {dy}}
\ee
As discussed above, this universality has immediate experimental
consequences which can be directly tested.

The second is a strong universality which is meant in a statistical
mechanical sense.  At first sight it appear a formal idea with little
relation to experiment.  If it is however true, its consequences are
very powerful and far reaching.  What we shall mean by strong
universality is that the effective action which describes small x
distribution function is critical and at a fixed point of some
renormalization group.  This means that the behavior of correlation
functions is given by universal critical exponents, and these universal
critical exponents depend only on general properties of the theory such
as the symmetries and dimensionality.

Since the correlation functions determine the physics, this statement
says that the physics is not determined by the details of the
interactions, only by very general properties of the underlying theory!

We can see how a renormalization group arises. The
effective action which we shall develop in later
lectures is valid only for gluons with rapidity less 
than $\eta_0$.\cite{mv}-\cite{mklw} where $\eta$ is the coordinate space 
rapidity.
Those at larger rapidity have been integrated out of the theory and
appear only as recoilless sources of color charge for $\eta_0 \le
\eta \le y_{proj}$.

The way a renormalization group is generated is by integrating out the
gluon degrees of freedom in the range $\eta_1 \le \eta \le \eta_o$,
to generate a new effective action for rapidity $\eta \le \eta_1$.  We
can show that this results in new source strength of color charge,
now in the range $\eta_1 \le \eta  \le y_{frag}$, and in a modification
of some of the coefficients of the effective action.

At high $Q^2$, the renormalization group analysis simplifies, and one
can show that in various limits reduces to the BFKL or DGLAP analysis.
The renormalization group equations at smaller $q^2$ become more
complicated, and have yet to be written in explicit form and evaluated.
This is in principle possible to do.\cite{ilm}

An essential ingredient in this analysis is the appearance of an action
and of a classical  gluon field.    We should expect the appearance of a
classical gluon field when the phase space density of gluons becomes
high.  Gluons are after all bosons, and when the phase space density is
large they should be described classically.  This provides the essential
difference between older renormalization group analysis which was
formulated in terms of a  distribution function.  High density and its
complications due to coherence require the introduction of a field.
The new renormalization group analysis is phrased in terms of the
effective action for the classical gluon field and jumps from an
ordinary equation to a functional equation.

The classical gluon fields which we shall find are the non-abelian
generalization of the Lienard-Wiechart potentials of electrodynamics.
If we use the coordinate space variable $x^-$ and realize that the
source for the gluons arise from much smaller $x^-$ than that at which
we make the measurements, since they arise from gluons of much higher
longitudinal momenta which are more Lorentz contracted, then the sources
can be imagined as arising from a $\delta (x^-)$.  The
Lienard-Wiechart potentials also are proportional to $\delta (x^-)$,
and so exist only in the sheet.  The fields are also transversely
polarized
\be
        B^i_a \perp E^i_a \perp \hat{z}
\ee

\subsection{Why an Effective Action?}

The effective action formalism which will be advocated in the next
lectures is very powerful.  It is used to compute a gluon effective
field.  This field can be related to the wavefunction of the hadron.

This field allows one to generalize it from the original description of a
single hadron, to collisions of hadrons and also to diffractive
processes.  We shall see that the effective action formulation is
incredibly powerful.

The careful student might at this point be very worried:  How has the
problem been in any way simplified?  We have just introduced a source,
and to specify the field one has to specify the source.  Moreover, such
a specification is gauge dependent.  What happens is amusing:  We
integrate over the all color orientations of the source.  Gauge
invariance is restored by the integration.  The theory becomes specified
by the local density of color charge squared.  This is a gauge invariant
quantity.   It can be related to the gluon distribution functions, and is
determined by the renormalization group equations.

\section{Small x Physics and Classical Fields}

In this lecture, I will develop the theory of small x physics
using classical gluon fields.  The reason why classical gluon fields are
important is because at small x the phase space density of gluons,
\be
	{{dN} \over {d^3xd^3p}} >> 1
\ee
When the occupation number per state becomes large, the quantum dynamics
becomes classical.

Before developing the classical theory, it is useful to review some
features of QCD, in particular, the light cone formulation.

\subsection{Light Cone Gauge QCD}

In QCD we have a vector field $A^\mu_a$.  This can be decomposed into
longitudinal and transverse parts as
\be
        A^\pm_a = {1 \over \sqrt{2}} (A^0_a \pm A^z)
\ee
and the transverse as lying in the tow dimensional plane orthogonal to
the beam z axis.  Light cone gauge is
\be
        A^+_a = 0
\ee

In this gauge, the equation of motion
\be
        D_\mu F^{\mu \nu} = 0
\ee
is for the $+$ component
\be
        D_iF^{i+} - D^+F^{-+} = 0
\ee
which allows one to compute $A^-$ in terms of $A^i$ as
\be
        A^- = {1 \over \partial^{+2}} D^i \partial^+ A^i
\ee
This equation says that we can express the longitudinal field entirely
in terms of the transverse degrees of freedom which are specified by
the transverse fields entirely and explicitly.  These degrees of freedom
correspond to the two polarization states of the gluons.

We therefore have
\be
        A^i_a (x) = \int_{p^+ > 0} {{d^3p} \over {(2\pi)^3 2p^+}} \left(
e^{ipx} a^i_a(p) + e^{-ipx} a^{i\dagger}_a (p)\right)
\ee
where
\be
        [a^i_a (p), a^{\j\dagger}_b(q)] = 2p^+ \delta_{ab} \delta^{ij} (2\pi)^3
\delta^{(3)} (p - q)
\ee
where the commutator is at equal light cone time $x^+$.

\subsection{Distribution Functions}

We would like to explore some hadronic properties using light cone field
operators.  For example, suppose we have a hadron and ask what is the
gluon content of that hadron.  Then we would compute
\be
        {{dN_{gluon}} \over {d^3p}} = <h\mid a^\dagger (p) a(p) \mid h>
\ee
The quark distribution for quarks of flavor $i$ (for the sum of quarks
and antiquarks) would be given in terms of creation and annihilation
operators for quarks as
\be
        {{dN_i} \over {d^3p}} = <h \mid \{b_i^\dagger (p) b_i(p) + d_i^\dagger
(p) d_i(p) \} \mid  h>
\ee
where $b$ corresponds to quarks and $d$ to antiquarks.
The creation and annihilation operators for quarks and gluons can be
related to the quark coordinate space field operators by techniques
similar to those above.  The interested student should read the notes of
Venugopalan for details.\cite{v}

How would we begin computing such distribution functions?  We will
start with the example of a large nucleus, as this makes some issues
conceptually simpler.  We will then generalize to hadrons, where we
shall see that the ideas presented here have a generalization.

For a large nucleus, we assume that the gluon distribution which we
shall try to compute has longitudinal momentum soft compared to that of
the valence quarks.  Valence quarks have longitudinal momentum of the
order of that of the nucleus, so this requirement is only that $x << 1$
In fact, we will require that $ x << A^{-1/3}$.  This is the requirement
that in the frame where the longitudinal momentum of the gluons is zero,
the nucleus has a Lorentz contracted size much less than the wavelength
associated with the gluons transverse momentum $\lambda \sim 1/p_T$.
This is the requirement that the gluon resolve the nucleus as a whole
and is insensitive to the details of the nuclear structure (spatial
distribution of valence quarks within the nucleus).
We are interested in the region where the overlap between the
quark and gluon rapidity distribution is small.

If the gluon phase space density is very large, the quantum gluons of
the nucleus may be treated as classical fields.  This may be true if
\be
        \Lambda^2 = {1 \over {\pi R^2}} {{dN_{glue}} \over {dy}}
\ee
satisfies $\Lambda^2 >> \Lambda^2_{QCD}$.  Certainly if the typical
$p_T$ of the gluons was of order $\Lambda_{QCD}$ this would be true since
the gluons would then be closely packed together.  We shall see that
this is true in fact for gluons with $p_T << \Lambda$ which for high
gluons density can become very large.

If the valence quarks have a longitudinal momentum much larger than the
typical gluon momentum then their typical interactions with these gluons
should be characterized by the soft momentum scale (otherwise the soft
gluons would not remain soft).  In an emission of a gluon, the emitted
gluon has momentum very small compared to the valence quark longitudinal
momentum.  Its velocity therefore is barely changed by this emission.
The quarks are therefore recoilless sources of color charge for the
gluon classical field.

The valence quarks
are sources for the gluons and sit on a sheet of thickness infinitesimal
compared to the typical wavelength associated with the gluon field.
We shall further assume for simplicity that the transverse distribution
of charge is uniform.  (In fact, a better approximation is that the
distribution of charge is slowly varying on the gluon wavelength scale
of interest.  This case can be computed directly from knowledge of the
uniform transverse distribution case,  The transverse size scale of
variation is the nuclear radius, which is much larger than a fermi, so
that this criteria is satisfied so long as $\Lambda >> \Lambda_{QCD}$,
and the typical transverse momentum is of order $\Lambda$.)

In a head on picture of the nucleus, it is essentially a flat sheet.  We 
are interested in resolving the charge distribution on a transverse
scale size of order $ \Delta x$.  In order that
we can use weak coupling methods, we require that the transverse size be
$\Delta x << 1 Fm.$

We can also resolve the structure in the longitudinal direction.  Here we
use spatial rapidity which has the effect of spreading out the thin
longitudinal extent of the nucleus, $\Delta z$ into a finite range of
rapidity.  If we draw a tube through the nucleus in the longitudinal
direction, then even for a very thin tube $\Delta x $ small, if we are at
high enough energy corresponding to a large rapidity, the tube will
intersect many quarks and gluons.   
The physical extent 
in longitudinal spatial coordinate is of course small.
If we require that $\Delta x << 1 Fm$ and require that a tube intersects
a quark or gluon from some hadron, then typically, that hadron will be
far separated from any other hadron which has one of its quarks or
gluons within the tube.  The quarks and gluons within the tube are
therefore uncorrelated in color.  If we further require that the tube be
large enough so that many quarks and gluons are in each tube, a
situation always possible if we have large enough energy or a large
enough nucleus, then the color total color charge associated with these
quarks and gluons will be in a high dimensional representation, and can
be treated classically.  To see this, recall that in a high dimensional
representation, $ Q^2 >> Q \sim [Q,Q]$ so that commutators may be
ignored.

For partons at the x values where we wish to compute the gluon
distribution function, the higher rapidity sources sit on a sheet of
infinitesimal thickness.  The problem we must solve is to compute the
typical correlation functions which give distribution functions for
stochastic sources sitting on a sheet of infinite transverse extent
traveling at the speed of light.  Mathematically, this problem is
similar to spin glass problems in condensed matter physics.  Such
problems typically have entirely nontrivial renormalization groups, and
we will see that this is the case for our problem.

To understand a little better how the sources are distributed, it is
useful to go inside the sheet.  This is accomplished by introducing a
space-time rapidity variable.  Let us assume that the rapidity of the
projectile is $y_{proj}$ and its longitudinal momentum is $P^+$, we
define
\be
        \eta = y_{proj} - ln(P^+_{proj} x^-)
\ee
We define the momentum space rapidity as
\be
        y = y_{proj} - ln(P^+_{proj}/p^+ )
\ee
Here $x^-$ is the coordinate of a source and $p^+$ its momentum.    The
previous definition of momentum space rapidity for produced particles
was
\be
        y = {1 \over 2} ln(p^+/p^-) = ln(p^+/m_t)
\ee
where $m_t = \sqrt{p_t^2 + m^2}$.   This last expression is valid for
particles on mass shell, whereas the other definition of momentum space
rapidity works for constituents of the hadron wavefunction, which are of
course not on mass shell.  These definitions are equal within about a
unit of rapidity for typical values of transverse momenta.  In the
wavefunction, the uncertainty principle relation gives $x^- p^+ \sim 1$,
so that we see the space time rapidity is up to about a unit of rapidity
the same as both momentum space values.  We will therefore use these
rapidities interchangeably in what follows.

We can now implement our static sources of charge by assuming the theory
is defined by an ensemble of such charges:
\be
        Z = \int [d\rho ] exp\left\{- {1 \over 2} \int_y^{y_{proj}} dy^\prime
d^2x_t {1 \over {\mu^2(y^\prime)}} \rho^2(y^\prime,x_t) \right\}
\ee
This ensemble gives
\be
        <\rho^a(y,x_T) \rho^b(y^\prime, x_t^\prime) > = \mu^2(y) \delta^{ab}
\delta(y-y^\prime) \delta^{(2)} (x_t - x_t^\prime)
\ee
The parameter $\mu^2$ therefore has the interpretation of a charge
squared per unit rapidity
\be
        \mu^2(y) = {1 \over {N_c^2-1}} {1 \over {\pi R^2}}{{dQ^2(y)}
 \over {dy}}
\ee
The total charge squared is
\be
        \chi  (y) = \int_y^{y_{proj}} dy^\prime \mu^2 (y^\prime)
\ee
Since the sources are individual quarks and gluons, this can also be
related directly to the total number of gluons and quarks contributing
to the source as\cite{gm}
\be
        \chi (y) = {1 \over {\pi R^2}} \left( {{N_g} \over {2N_c}} + {{N_cN_q}
\over {N_c^2-1}} \right) \int_x^1 dx G(x)
\ee
The factors above arise from computing the color charge squared of a
singe quark or a single gluon.

We may now construct the non-abelian Lienard-Wiechart potentials
generated by this distribution of sources.  We must solve the equation
\be
        D_\mu F^{\mu \nu} = g^2 \delta^{\nu +} \rho (y, x_t)
\ee
To solve this equation, we look for a solution of the form
\be
        A^\pm = 0
\ee
and
\be
        A^i = {1 \over i} U(y,x_t) \nabla^i U^\dagger (y,x_T)
\ee
We could equally well solve this equation in a gauge where $A^+$ is
nonzero and all other components vanish.  The student should check that
the gauge transformation induced by $U$ gives
\be
        \overline A^+ = { 1 \over i} U^\dagger \partial^+ U
\ee
In this non-lightcone gauge, the field equations simplify and we get
\be
        \nabla_t^2 \overline A^+ = g^2 \rho (y,x_t)
\ee
Solving for $U$ gives
\be
        U = exp\left\{i \int_y^{y_{proj}} dy^\prime {1 \over \nabla_t^2}
\rho(y^\prime, x_t) \right\}
\ee
We have therefore constructed an explicit expression for the light cone
field $A^i$ in terms of the source $\rho$ for arbitrary $\rho$!  The
system is integrable.

The structure of the fields strengths $E$ and $B$ which follow from this
field strength is now easy to understand.  The only nonzero longitudinal
derivative is $\partial^+$.  The field strength $F_{ij}$ where both $ij$
are transverse vanishes since the filed looks like a pure gauge
transformation in the two dimensional space.  It is in fact a pure gauge
everywhere but in the sheet where the charge sits.  Therefore the only
nonvanishing field strength is $F^{i+}$.  This gives $E \perp B \perp
\vec{z} $, that is the fields are transversely polarized.  This is the
precise analog of the Lienard-Wiechart potentials of electrodynamics.

\subsection{The Action and Renormalization Group}

The effective action for the theory we have described must be gauge
invariant and properly describe the dynamics in the presence of external
sources, up to an overall gauge transformation which is constant in $x^\pm$.
(The lack of precise gauge invariance arises from the the desire to define the
intrinsic transverse momentum of gluon distribution functions.  Although when used
in computations of gauge dependent quantities, the gauge dependence disappears,
it is useful to not have the full gauge invariance in intermediate steps of computation.)  
The student should verify, that consistency with the
Yang-Mills equations in the presence of an external source requires that
\be
        D_\mu J^\mu = 0
\ee
For a source of the type we have here,
\be
        J^\mu_a = \delta^{\mu a} \delta (x^-) \rho^a(x_t)
\ee
This is an approximation we make when we describe the
source on scales much larger than the spatial extent of the
source.  To properly regularize the delta function, we need
to spread the source out in $x^-$ as was done in the
previous section.
We find that the action is
\be
        S = - {1 \over 4} \int d^4x F_{\mu \nu}^aF_a^{\mu \nu}
+ {i \over N_c} \int d^2x_t dx^- \delta (x^-) \rho^a(x_t) tr T^a
exp\left\{ i \int _{-\infty}^\infty dx^+ T \cdot A^-(x) \right\}
\ee
In this equation, the matrix $T$ is in the adjoint representation of the
gauge group.  This is required for reality of the action.  The student
should minimize this action to get the Yang-Mills equations, identify
the current, and show that the current is covariantly conserved.

This action is gauge invariant under gauge transformations which are
required to be periodic in the time $x^+$.  This is a consequence of the
gauge invariance of the measure of integration over the sources $\rho$. 
This will be taken as a boundary condition upon the theory.   In general
if we had not integrated over sources, one could not define a gauge
invariant source, as gauge rotations would change the definition of the
source.  Here because the source is integrated over in a gauge invariant
way, the problem does not arise.

In the most general gauge invariant theory which we can write down
is generated from
\be
        Z = \int [d \rho ] e^{-F[\rho ]} \int [dA] e^{iS[A,\rho ]}
\ee
This is a generalization of the Gaussian ansatz described in the
previous lecture.  It allows for a slightly more complicated structure
of stochastic variation of the sources.  The Gaussian ansatz can be
shown to be valid when the evaluating structure functions at large
transverse momenta.
\be
        F_{Gaussian} [\rho ] = { 1 \over {2 \chi }} \int d^2 x_t \rho^2 (x_t)
\ee

This theory is a slight modification of what we described in the first
lecture.  We have here assumed that our theory is an effective theory
valid only in a limited range of rapidity much less than the rapidity 
of the source.  In this restricted range,
the structure of the source in rapidity cannot be important, and
therefore we couple only to the total charge seen at the rapidity of
interest.   The local charge density as a function of rapidity is replaced
by the total charge at rapidities greater than that at which we measure 
the field.  The scale of fluctuation of the source is instead of the local
charge squared per unit are per unit rapidity $\mu^2 (y)$ becomes replaces by
the scale of fluctuation of the total charge.  The student should prove that
\be
     \chi = \int_y^{y_{proj}} dy^\prime \mu^2 (y^\prime)
\ee

To fully determine $F$ in the above equation demands a full solution of
the renormalization group equations of the theory.  This has yet to be
done.  We shall confine our attention in most of the analysis below to
the Gaussian ansatz.  It is remarkable that within this simple ansatz
for $F$, most of the general feature which a full treatment should
generate, such as proper unitary behavior of deep inelastic scattering,
arises in a natural way.

Let us turn our attention for the moment to the gluon distribution
function
\be
        (2\pi )^3 2p^+ {{dN} \over {d^3p}} = <a^\dagger (p) a(p) >
\ee
Using the results of the previous lecture, you should prove that
\be
        {{dN} \over {d^3p}} = {{2p^+} \over {(2\pi )^3}} \sum_{i,a} D^{ii}_{aa}
(p, -p)
\ee
Here $a$ is a color index and $i$ is a transverse index associated with
the gluon field.  $D$ is the gluon propagator in the external field
\be
        D^{\mu \nu}_{ab} (p,q) = <A^\mu_a(p) A^\nu_b(q)>
\ee
(In field theory language, $D$ is the propagator in the external field
including both connected and disconnected pieces.)

In lowest order, the $A$ in the expression for $D$ is simply the
external Lienard-Wiechart potential.  Recall that
\be
        A^i = -i U \nabla^i U^\dagger
\ee
where the $U$'s are the explicit functions of the source $\rho$ computed
in the previous lecture. 

The expectation value above may be computed using the Gaussian weight
function.  We need the propagator $1/\nabla_t^4$ to perform this
calculation.  The student should try to do this computation (it is
equivalent to normal ordering exponentials), and if it is too hard, refer
to the paper of Jalilian-Marian et. al, where it is worked out in some
detail.\cite{mkmw}  The result is
\be
        {{dN} \over {d^2p_T}} \sim \int d^2z e^{-ip_T z_T} ~ {{4(N_c^2-1)} \over {N_c z_T^2}}
\theta(1-z_t \Lambda_{QCD})    \left\{ 1 - (z_t^2 \Lambda_{QCD}^2)^{2\pi
\alpha_s^2 \chi z_t^2} \right\}
\ee

At $p_t >> \Lambda$ where $\Lambda = \alpha_s \sqrt{\chi}$, the
\be
        {1 \over {\pi R^2}} {{dN} \over {dy d^2p_t}} \sim \Lambda^2/ 
\alpha_s p_t^2
\ee
This is because this part of the distribution can be thought of as
arising from bremstrahlung from the sources of the gluon field at high
rapidities.  As $p_t \le \Lambda$, the gluon distribution function
goes to a slowly varying function of $p_T$, which in the Gaussian ansatz
is
\be
        {1 \over {\pi R^2}} {{dN} \over {dy d^2p_t}} \sim {1 \over \alpha_s}
ln(\Lambda /p_t)
\ee
In general, we expect saturation in this region, for arguments which
will be presented in the next few pages.  This means we get a slowly
varying function of $p_T$, which by dimensional arguments, is therefore
a slowly varying function of $\Lambda $

The only place that any non-trivial rapidity dependence enters the
problem is through $\Lambda$.  This dependence can be found from the
renormalization group analysis of the effective action.  Therefore in
the bremstrahlung region, the dependence on $\Lambda^2 $ is linear, and in
the saturation region it is very weak.

Let us assume that when we go beyond the Gaussian ansatz, the general
features of saturation presented above remain.  We will show that this
results in a reasonable physical picture for small x physics, and solve
the unitarity puzzle outlined in the first lecture.  First consider the
gluon distribution function.  The total number of gluons which can be
measured at some scale size larger than a resolution size scale $1/Q^2$
is
\be
        xG(x,Q^2) = \int_0^{Q^2} {{d^2p_T} \over {(2\pi)^2}} {{dN} \over
{dyd^2p_T}}
\ee

For $Q^2 >> \Lambda^2 $, the integral is dominated by the
bremstrahlung tail, and $G \sim R^2 \Lambda^2 $.  (The gluon 
distribution is proportional to the charge per unit area times the area.)
In our random walk
scenario, the effective charge squared must be proportional to the
length of the random walk, $R$, so that $G \sim R^3$ which for a nucleus is  
proportional to $A^{1/3}$ .  This is what we expect
at large $Q^2$, except the reasoning is a little different than usual. 
The standard argument would have been that at large $Q^2$, the degrees
of freedom in a nucleus for example should act incoherently, and we
should have the gluon distribution functions proportional to $A$.  Here
we have random fields generating precisely the same $A$ dependence!

When $Q^2 \le \Lambda^2 $, the integral is dominated by the
saturation region.  Here $G \sim R^2 Q^2$, and the gluons can be thought
of as arising from the surface of the hadron.  Again this is consistent
with what is expected from phenomenology.  The gluons are so soft that
they cannot see the entire hadron, only its surface.

The effect of saturation for unitarization of deep inelastic scattering
can also be easily understood.  Suppose we are at some fixed $Q^2 >>
\Lambda^2(y)$.  As $y$ decreases corresponding to going to smaller
$x$,, the gluon distribution function increases as $\Lambda^2 (y)$.  When $x$
becomes so small that we get into the saturation region, the linear
$\Lambda $ dependence is weakened (in the Gaussian ansatz it is logarithmic
in $\Lambda $), and the structure function stops growing.  We expect that
the number of gluons of size smaller than our resolution scale have
ceased to grow, and the cross section should become slowly varying.
This is in spite of the fact that $\Lambda $ continues to grow!

The physics is again simple to understand:  At small x we are indeed
adding more and more gluons to the hadron wavefunction.  These gluons
are smaller as their inverse size scales as gluon density,
\be
        1/r^2 \sim {1 \over {\pi R^2}} {{dN} \over {dy}}
\ee
They do not contribute to a fixed $Q^2$ cross section when they become
smaller than the resolution size scale.  What saves unitarity is $p_t$
broadening.

\subsection{Deep Inelastic Scattering}

In the previous discussion, we were concerned with the gluon
distribution function.  In deep inelastic scattering, we measure the
quark distribution functions.  How are these related?

In deep inelastic scattering we measure
\be
        W^{mu \nu } & = & { 1 \over {2\pi}} Im \int
d^4x e^{iqx} <P_{had} \mid T(J^\mu (x) J^\nu (0) \mid P_{had} >
\nonumber \\
   & = & {1 \over M_{had}} \left\{ - \left(  g^{\mu \nu} -{{q^\mu q^\nu}
\over q^2} \right) F_1 + \right. \nonumber \\
& & \left. \left( p^\mu - q^\mu {{p \cdot q} \over q^2}
\right) \left( p^\nu - q^\nu {{p \cdot q} \over q^2} \right) {1 \over {p
\cdot q}} F_2 \right\}
\ee

We must be able to compute the imaginary part of the current-current
correlation function.     This is simply vacuum polarization in the
presence of the Lienard-Wiechart potentials.  In fact, it is
straightforward to compute the propagators in these 
background fields.\cite{mv1} 
The reason why it is simple is because the background field is basically
gauge transformations in two regions of spaces separated by a surface of
discontinuity. 

This propagator may be then used to compute the vacuum polarization. 
The computation can be done so far as to relate explicitly the structure
functions to correlation functions of exponentials involving $\rho $ and
expressions very similar to those for the gluon distribution function
result.  These results are currently being applied to the study of deep
inelastic scattering to see whether the effects of saturation might be
seen experimentally.

\subsection{Renormalization Group and How It Works}

In the above computation of the gluon propagator, one used the classical
effective action to compute the Lienard-Wiechart potentials.  These
classical fields were then used to compute the propogator.  What about
the quantum corrections?

When the quantum corrections are computed, one gets a correction to the
classical field of order $\alpha_s ln(x_{cutoff}/x)$, where $x$ is the
value for gluon distribution function, and $x_{cutoff}$ is the maximum
$x$ appropriate for the effective action.  Although the coupling is
small, the logarithm can become big if we go to $x$ values far below the
cutoff.  The overall theory should remain valid however, the problem is
that the classical approximation is no longer good.

The way to fix this up is to use the renormalization group.  Suppose we
want a new effective theory below $x_{new}$, and we require that
$x_{new} << x_{cutoff}$ but that $\alpha_s ln(x_{cutoff}/x_{new}) << 1$.
This means that the quantum corrections are small in the range
$x_{new} << x << x_{cutoff}$, and can be handled perturbatively.
We proceed by integrating out the degrees of freedom in this
intermediate range of x, to generate a new effective theory in the
region $x << x_{new}$.

In this process, the one can show that the only thing that changes in
the effective action formalism is the weight function $F[\rho ]$ for
fluctuations in the color source .  Effectively, we are changing what
we call source for the gluon field, and trading it for what 
we call the gluon field.
This renormalization group procedure is that of Wilson-Kadanoff.  One
can derive the form of the renormalization group equations and they can
be written explicitly in the low gluon density region.  One can prove
that here the function $F$ is Gaussian, and that in appropriate
kinematic limits, one reproduces the BFKL equation, the DGLAP equation
and their non-linear generalizations to first order in the
non-linearities.  Work is currently in progress to get the explicit form
of these equations in the high density region, and to solve them.
 
If we are in the low density region, we evolve in $x$ by the
BFKL equation and in $Q^2$ by the DGLAP equation.   Hopefully a full
solution to the problem will lead to an effective action which is at a
fixed point of the renormalization group.  In this case, at very small
x, the form for $F$ will simplify, and all correlation functions will
have universal critical exponents.  If this is true, the dynamics of
high energy scattering for all hadrons becomes the same, and presumably
has a simple structure.

\section{Acknowledgments}

I thank my colleagues Alejandro Ayala-Mercado, Miklos Gyulassy,
Yuri Kovchegov, Alex Kovner, Jamal Jalilian-Marian, Andrei Leonidov,
Raju Venugopalan and Heribert Weigert with whom the ideas presented
in this talk were developed.  This work was supported under Department
of
Energy grants in high energy and nuclear physics DOE-FG02-93ER-40764
and DOE-FG02-87-ER-40328.

\end{document}